\documentclass[aps,pra,twocolumn,showpacs,amsmath,amssymb,10pt]{revtex4-1}
\usepackage{graphicx}
\usepackage{bm}
\usepackage{slashed}
\usepackage{color}

\newcommand{\dd}{\mathrm{d}}         

\begin{document}
\title{Electron dressed mass in short laser pulses and its measurement}
\thanks{Work presented during the International Laser Physics Workshop 2015, Shanghai, China}
\author{F. Cajiao V\'elez$^1$}
\author{K. Krajewska$^{1,2}$}
\email[E-mail address:\;]{Katarzyna.Krajewska@fuw.edu.pl}
\author{J. Z. Kami\'nski$^1$}
\affiliation{$^1$Institute of Theoretical Physics, Faculty of Physics, University of Warsaw, Pasteura 5,
02-093 Warszawa, Poland \\
$^2$Department of Physics and Astronomy, University of Nebraska, Lincoln, Nebraska 68588-0299, USA
}

\date{\today}
\begin{abstract}
The concept of the electron mass dressing by a powerful laser pulse is discussed. It is shown, by considering the coherent frequency combs generated out of the Compton radiation,
how the electron dressed mass can be determined experimentally. This also opens a possibility to measure properties of extremely intense pulses for which the previously developed methods, 
working at moderate intensities, are not applicable. Namely, one can determine these properties from the properties of coherent Compton radiation.
\end{abstract}
\maketitle

\section{Introduction}
\label{sec:motive}

The concept of the electron mass dressing by a laser field is almost as old as the laser itself. It was V.~I.~Ritus, A.~I.~Nikishov and their collaborators 
(see, e.g., \cite{RitusNikishov,Ritus,RitusNikishov2} and references therein) who thoroughly investigated the importance of this concept by considering 
fundamental processes of quantum electrodynamics (QED), modified or induced by an action of intense electromagnetic plane wave (for recent reviews see, 
e.g., Refs.~\cite{EKK2009,Rev1,PMHK2012,Ras,Narozhny}). The prominent example of such a process is the Compton scattering in which an electron, 
interacting with a laser field, emits a nonlaser photon. Its classical analogue is the so-called Thomson scattering in which an electron, accelerated or decelerated while interacting 
with a laser field, emits an electromagnetic radiation.

If an electron interacts with a monochromatic electromagnetic plane wave of the frequency $\omega_{\mathrm{L}}$ and of the time-averaged intensity $I$, 
the energy spectrum of Compton photons is defined by the so-called Klein-Nishina formula~\cite{KleinNishina,BrownKibble},
\begin{equation}
\omega_{\bm{K},N}=\frac{N\omega_{\mathrm{L}}}{\frac{p_{\mathrm{i}}\cdot n_{\bm{K}}}{p_{\mathrm{i}}\cdot n}
+\frac{U}{c}\frac{n\cdot n_{\bm{K}}}{p_{\mathrm{i}}\cdot n}+\frac{N\omega_{\mathrm{L}}}{\omega_{\mathrm{cut}}}},
\label{comp1}
\end{equation}
with positive integers $N$. In this formula, the symbol $a\cdot b$ means the relativistic scalar product of four-vectors, $a\cdot b=a^0b^0-\bm{a}\cdot\bm{b}$, with the four-vector $a=(a^0,\bm{a})$ 
and similarly for $b$, where $\bm{a}\cdot\bm{b}$ is the three-dimensional scalar product. Here, it is assumed that the plane wave propagates in the space direction $\bm{n}$, meaning that $n=(1,\bm{n})$ and $n\cdot n=0$. Similarly, the Compton photon is generated in the 
direction $\bm{n}_{\bm{K}}$ [$n_{\bm{K}}=(1,\bm{n}_{\bm{K}})$ and $n_{\bm{K}}\cdot n_{\bm{K}}=0$] with the four-momentum $K_N=(\omega_{\bm{K},N}/c)n_{\bm{K}}$, and 
$p_{\mathrm{i}}=(p^0_{\mathrm{i}},\bm{p}_{\mathrm{i}})$ is the electron initial momentum on the free-electron mass shell, $p_{\mathrm{i}}\cdot p_{\mathrm{i}}=(m_{\mathrm{e}}c)^2$. The so-called ponderomotive energy of electrons in the laser field equals
\begin{equation}
U=\frac{1}{4}\mu^2\frac{(m_{\mathrm{e}}c^2)^2}{cp_{\mathrm{i}}\cdot n}.
\label{comp1a}
\end{equation}
Here, 
\begin{equation}
\mu=\frac{|e|\mathcal{E}_0}{m_{\mathrm{e}}c\omega_{\mathrm{L}}},
\label{comp1b}
\end{equation}
where the electric field amplitude of the plane wave, $\mathcal{E}_0$, is related to the time-averaged intensity of the laser field such that ($\varepsilon_0$ means the vacuum permittivity)
\begin{equation}
I=\frac{1}{2}\varepsilon_0 c\mathcal{E}_0^2.
\label{comp1c}
\end{equation}
Moreover,
\begin{equation}
\omega_{\mathrm{cut}}=\frac{c}{\hbar}\frac{p_{\mathrm{i}}\cdot n}{n\cdot n_{\bm{K}}}
\label{comp2}
\end{equation}
defines the maximum value of $\omega_{\bm{K},N}$ (i.e., $\omega_{\bm{K},N}\rightarrow\omega_{\mathrm{cut}}$ when $N\rightarrow\infty$), 
and it represents the only parameter that contains the Planck constant $\hbar$. Note that, in the limit $\hbar\rightarrow 0$, we recover the classical 
Thomson formula for frequencies of generated radiation,
\begin{equation}
\omega^{\mathrm{Th}}_{\bm{K},N}=\frac{N\omega_{\mathrm{L}}}{\frac{p_{\mathrm{i}}\cdot n_{\bm{K}}}{p_{\mathrm{i}}\cdot n}+\frac{U}{c}\frac{n\cdot n_{\bm{K}}}{p_{\mathrm{i}}\cdot n}},
\label{comp3}
\end{equation}
which are not limited from above as, in the classical limit, $\omega_{\mathrm{cut}}\rightarrow \infty$. Hence, one can find the scaling law,
\begin{equation}
\omega^{\mathrm{Th}}_{\bm{K},N}=\frac{\omega_{\bm{K},N}}{1-\frac{\omega_{\bm{K},N}}{\omega_{\mathrm{cut}}}},
\label{comp4}
\end{equation}
which relates the Compton and Thomson frequencies. The consequences of this scaling law for the plane waves and long laser pulses have been 
studied in~\cite{SKscale1,SKscale2,SKscale3}, whereas its extension to arbitrary short laser pulses with the discussion of the electron spin and laser pulse polarization effects has been investigated in \cite{KKscaling}. In our further analysis we use units in which $\hbar=1$.

It follows from the Klein-Nishina formula~\eqref{comp1} that, for a given kinematics of the Compton scattering, the measurement of positions of two consecutive peaks in the Compton spectrum, 
$\omega_{\bm{K},N}$ and $\omega_{\bm{K},N+1}$, allows one to determine the ponderomotive energy $U$. Indeed, after some algebra, we arrive at the expression, 
\begin{align}
\frac{p_{\mathrm{i}}\cdot n_{\bm{K}}}{p_{\mathrm{i}}\cdot n}+\frac{U}{c}\frac{n\cdot n_{\bm{K}}}{p_{\mathrm{i}}\cdot n}&=\frac{\omega_{\mathrm{L}}}{\omega_{\bm{K},N+1}-\omega_{\bm{K},N}}\nonumber\\
& \times\Bigl(1-\frac{\omega_{\bm{K},N+1}}{\omega_{\mathrm{cut}}}\Bigr) \Bigl(1-\frac{\omega_{\bm{K},N}}{\omega_{\mathrm{cut}}}\Bigr).
\label{comp5}
\end{align}
Note that the right-hand-side of this expression has to be independent of $N$. This means that the proposed procedure of measuring $U$ does not require the knowledge of the order of the Compton 
frequency, $N$, but only positions of arbitrary chosen two adjacent peaks. In general, one could select peaks in the spectrum corresponding to frequencies $\omega_{\bm{K},N}$ and 
$\omega_{\bm{K},N+M}$, $M=1,2,\ldots$, and arrive at the similar conclusion.

In order to relate the ponderomotive energy $U$ to the electron mass dressing let us recall that the Volkov solution of the Dirac equation \cite{Volkov} for an electron with the four-momentum 
$p=(p^0,\bm{p})$, $p\cdot p=m^2_{\mathrm{e}}c^2$, is
\begin{equation}
\psi^{(+)}_{\bm{p}\lambda}(x)=\sqrt{\frac{m_{\mathrm{e}}c^2}{VE_{\bm{p}}}}\Bigl[ 1-\frac{e}{2k\cdot p}\slashed{A}(k\cdot x)\slashed{k}\Bigr] u^{(+)}_{\bm{p}\lambda}\mathrm{e}^{-\mathrm{i}S^{(+)}_p(x)},
\label{comp6}
\end{equation}
where
\begin{equation}
S^{(+)}_p(x)=p\cdot x+\int_0^{k\cdot x}\Bigl[e\frac{A(\phi)\cdot p}{k\cdot p}-e^2\frac{A^2(\phi)}{2k\cdot p}\Bigr]\mathrm{d}\phi,
\label{comp7}
\end{equation}
$E_{\bm{p}}=cp^0$, and $u^{(+)}_{\bm{p}\lambda}$ is the free-electron bispinor normalized such that 
$\bar{u}^{(+)}_{\bm{p}\lambda}u^{(+)}_{\bm{p}\lambda'}=\delta_{\lambda\lambda'}$, with $\lambda=\pm$ labeling its spin degrees of freedom. 
Moreover, for the linearly polarized plane wave, the vector potential has the form,
\begin{equation}
A(\phi)=\varepsilon\frac{\mathcal{E}_0}{\omega_{\mathrm{L}}}\cos\phi,
\label{comp8}
\end{equation}
where $\varepsilon$ is the polarization four-vector such that $\varepsilon=(0,{\bm{\varepsilon}})$ and $\varepsilon\cdot\varepsilon=-{\bm{\varepsilon}}\cdot{\bm{\varepsilon}}=-1$.
Note also that $k=(\omega_{\mathrm{L}}/c)(1,\bm{n})$, with the unit vector $\bm{n}$ defined above. Hence, the function $S^{(+)}_p(x)$ can be splitted into two parts
\begin{equation}
S^{(+)}_p(x)=\bar{p}\cdot x+ G_p(k\cdot x),
\label{comp9}
\end{equation}
where $G_p(\phi)$ is a periodic function of $\phi$ with the period of $2\pi$, whereas
\begin{equation}
\bar{p}=p+\frac{1}{4}(\mu m_{\mathrm{e}}c)^2\frac{k}{p\cdot k}
\label{comp10}
\end{equation}
is the so-called electron dressed momentum by the laser field. One can easily verify that
\begin{equation}
\bar{p}\cdot\bar{p}=(\bar{m}_{\mathrm{e}}c)^2,\quad \bar{m}_{\mathrm{e}}=m_{\mathrm{e}}\sqrt{1+\frac{1}{2}\mu^2}.
\label{comp11}
\end{equation}
This means that the electron momentum in the laser field is on the mass shell with the dressed mass $\bar{m}_{\mathrm{e}}$. By determining the ponderomotive energy $U$ and,
hence, the parameter $\mu$ [see, Eq.~\eqref{comp1a}], the electron dressed mass [Eq.~\eqref{comp11}] can be directly determined from the Compton spectrum.

The procedure described above allows one to determine the electron mass dressing in the presence of a monochromatic plane wave or, after some modifications, in the presence
of polychromatic plane waves with commensurate frequencies. In real experiments, however, one deals with finite-in-time and focused-in-space laser pulses. The requirement 
of space focusing can be soften by considering the head-on collision of the laser pulse with relativistic electrons. Under such conditions, 
in the electron reference frame, the laser focus is largely oblate. Thus, the laser pulse can be modeled as a plane-wave-fronted pulse which is finite in its propagation 
direction but it is infinite in the perpendicular directions~\cite{Lee2010}. For such laser pulses the following questions arise: 
\begin{itemize}
\item[(i)]
Is it still meaningful to consider the electron mass dressing by an arbitrary short laser pulse?
\item[(ii)]
If yes, can we measure the electron dressed mass for an arbitrary chosen laser pulse?
\end{itemize} 
Our purpose is to answer these two questions.

\section{Momentum dressing by short laser pulses}
\label{sec:compton}

Assume that the laser pulse lasts from $t=0$ up to $t=T_{\mathrm{p}}$ and vanishes beyond this time interval. Note that its time duration defines the fundamental frequency, 
$\omega=2\pi/T_{\mathrm{p}}$, which is the smallest frequency in the spectral decomposition of the pulse. If $k=(\omega/c)(1,\bm{n})$, then the 
electromagnetic vector potential has the form,
\begin{equation}
A(\phi)=A_0(\varepsilon_1f_1(\phi)+\varepsilon_2f_2(\phi)),\quad A_0=\frac{\mathcal{E}_0}{\omega},
\label{com1}
\end{equation}
where $\phi=k\cdot x$. Moreover, $\varepsilon_j=(0,\bm{\varepsilon}_j)$ for $j=1,2$ denote two linear polarizations of the laser field such that 
$\varepsilon_{j'}\cdot\varepsilon_j=-\delta_{j'j}$ and $k\cdot\varepsilon_j=0$. The two functions $f_j(\phi)$, $j=1,2$, which we call the shape functions, 
are arbitrary and sufficiently smooth functions that vanish for $\phi<0$ and $\phi>2\pi$. We also define their mean values as
\begin{equation}
\langle f_j^m\rangle=\frac{1}{2\pi}\int_0^{2\pi}\dd\phi \, [f_j(\phi)]^m, \quad j,m=1,2.
\label{com2}
\end{equation}
With these definitions, the function $S^{(+)}_p(x)$ [Eq.~\eqref{comp7}], can be decomposed as in~\eqref{comp9}.
This time, however, the electron dressed momentum becomes
\begin{align}
\bar{p}=p- & \mu m_{\mathrm{e}}c\Bigl(\frac{p\cdot\varepsilon_1}{p\cdot k}\langle f_1\rangle+\frac{p\cdot\varepsilon_2}{p\cdot k}\langle f_2\rangle\Bigr)k \nonumber \\ 
+ &
\frac{1}{2}(\mu m_{\mathrm{e}}c)^2\frac{\langle f^2_1\rangle+\langle f^2_2\rangle}{p\cdot k}\, k.
\label{com3}
\end{align}
Is it the most general form of the momentum dressing? The answer to this question depends on the QED process under considerations. 
As it follows from the analysis presented in~\cite{KKcompton}, for the Compton scattering, the scattering amplitude depends only on the difference between 
the final and initial dressed momenta, $\bar{p}_{\mathrm{f}}-\bar{p}_{\mathrm{i}}$. This means that the physical observables do not change if we redefine 
the laser field dressing such that 
\begin{equation}
\bar{p}_{\mathrm{f,i}}\rightarrow \bar{p}_{\mathrm{f,i}}+P,
\label{com4}
\end{equation}
where $P$ is an arbitrary four-vector. (Another transformation has to be introduced, for instance, for the Breit-Wheeler process, as it has been discussed in~\cite{KKbw}.)

In general, the four-vector $P$ can be represented as
\begin{equation}
P=g_1\varepsilon_1+g_2\varepsilon_2+g_0 k,
\label{com5}
\end{equation}
where, in the absence of the laser field, $g_j=0$ ($j=0,1,2$). It appears that the shifted dressed momentum~\eqref{com4} is on the mass-shell (i.e., $\bar{p}_{\mathrm{f,i}}\cdot\bar{p}_{\mathrm{f,i}}$ 
is independent of $p_{\mathrm{f,i}}$) for a particular choice of $g_j$ such that
\begin{equation}
g_1=\mu m_{\mathrm{e}}c\langle f_1\rangle, \quad g_2=\mu m_{\mathrm{e}}c\langle f_2\rangle, \quad g_0=0.
\label{volk8}
\end{equation}
In this case,
\begin{equation}
\bar{p}_{\mathrm{f,i}}\cdot\bar{p}_{\mathrm{f,i}}=(\bar{m}_{\mathrm{e}}c)^2=(m_{\mathrm{e}}c)^2\Bigl[1+\frac{2U_0}{m_{\mathrm{e}}c^2}\Bigr],
\label{volk9}
\end{equation}
where $\bar{m}_{\mathrm{e}}$ can be called the electron dressed mass in the laser pulse, whereas
\begin{equation}
U_0=\frac{1}{2}\mu^2  m_{\mathrm{e}}c^2[\langle f^2_1\rangle-\langle f_1\rangle^2+\langle f^2_2\rangle-\langle f_2\rangle^2] .
\label{volk10}
\end{equation}
Note that, for the monochromatic plane wave, when $\langle f_j\rangle=0$, $j=1,2$, and $\langle f^2_1\rangle+\langle f^2_2\rangle=1/2$, the last two equations
lead to the previous result~\eqref{comp11}. This answers our first question raised in Sec.~\ref{sec:motive}: {\it if one accepts the above definition of the momentum 
dressing~\eqref{com3} then the measurement of the electron dressed mass induced by a pulse is equivalent to determining its three characteristics}~\cite{KKgkn},
\begin{equation}
\mu \langle f_j\rangle,\, j=1,2, \quad \textrm{and}\quad \mu^2 [\langle f^2_1\rangle+\langle f^2_2\rangle].
\label{com6}
\end{equation}
Actually, this task belongs to a broader topic which is the diagnosis of powerful laser pulses. The point being that, for such pulses, the conventional diagnostic methods developed for moderately intense 
fields are not applicable. Therefore, one may determine the peak intensity of the laser pulse by studying the properties of radiation generated in either the laser-modified
recombination process~\cite{Krec1,Krec2,Carstenrec} or in the Thomson scattering~\cite{Har}. It seems that this goal can be 
realized even more effectively by analyzing the ionization spectrum of photoelectrons~\cite{Kalashnikov} (see, also Ref.~\cite{KKcombs2015}). Another possibility is to exploit 
the chirp of the laser pulse in the Thomson or Compton scattering \cite{Brad,Terzic,Seipt2015}. Moreover, the problem of the relative change of the electron mass dressing by two 
laser pulses of different shapes has been studied in~\cite{HHIM}. The latter, however, does not provide a method for a direct measurement of the electron dressed mass. 
The aim of the next Sections is to discuss how the electron dressed mass can be determined directly (and, from the theoretical point of view, very precisely even for extremely 
short one-cycle laser pulses) from the spectrum of radiation generated during the Compton scattering.

\section{Coherent frequency combs}

It is known that the coherent properties of radiation, which manifest themselves in interference and diffraction experiments, can lead to very precise measurements of 
physical quantities. This approach was initiated by T. Young and his famous double-slit experiment~\cite{Young}, which then triggered the development of other experimental techniques 
(like diffraction gratings). Among them, the frequency comb generation is of a particular interest. Generation of optical frequency combs~\cite{freqcombs1,freqcombs2} 
has become an invaluable experimental tool for precise measurements of optical frequencies, that also enabled the creation of optical atomic clocks~\cite{freqcombs3}. 
A particular realization of this technique, called the high-order harmonic generation~\cite{AgostiniMauro}, has led to the development of attoscience~\cite{FT,KrauszIvanov}. This newly growing field is 
related to the interaction of matter with attosecond pulses of coherent radiation. Also, it has been proposed that the interaction of modulated laser pulses with relativistic 
and nearly monoenergetic electron beams can create x- and $\gamma$-ray frequency combs in the keV or MeV region~\cite{KK1,KK2}. This may extend the applicability of frequency
comb generation to nuclear and elementary particle physics. The same phenomenon was studied for matter and anti-matter waves~\cite{KK1,KK3}.

It has been demonstrated for the Compton scattering stimulated by a finite train of identical pulses~\cite{KK1,KK2} that the following interference/diffraction formula,
which describes the probability amplitude of the process, $\mathcal{A}_{\sigma}(\omega_{\bm{K}},\lambda_{\mathrm{i}},\lambda_{\mathrm{f}})$, holds
\begin{align}
\mathcal{A}_{\sigma}(\omega_{\bm{K}},\lambda_{\mathrm{i}},\lambda_{\mathrm{f}})=&\exp\bigl[\mathrm{i}\Phi_{\sigma}(\omega_{\bm{K}},\lambda_{\mathrm{i}},\lambda_{\mathrm{f}}) \bigr]\frac{\sin(\pi\bar{Q}^+/k^0)}{\sin(\pi\bar{Q}^+/k^0N_{\mathrm{rep}})} \nonumber \\
\times & |\mathcal{A}^{(1)}_{\sigma}(\omega_{\bm{K}},\lambda_{\mathrm{i}},\lambda_{\mathrm{f}})|.
\label{comb1}
\end{align}
Here, $N_{\mathrm{rep}}$ is the number of laser pulses in the train, $\mathcal{A}^{(1)}_{\sigma}(\omega_{\bm{K}},\lambda_{\mathrm{i}},\lambda_{\mathrm{f}})$ is the probability amplitude 
of Compton scattering by a single pulse from the train, and $\Phi_{\sigma}(\omega_{\bm{K}},\lambda_{\mathrm{i}},\lambda_{\mathrm{f}})$ is the global phase. Moreover, $\sigma$ labels the polarization of the 
Compton photon, and $\lambda_{\mathrm{i}}$ and $\lambda_{\mathrm{f}}$ define the spin degrees of freedom of the initial and final electron states, respectively. 
The coherent enhancement of the Compton amplitude $\mathcal{A}_{\sigma}(\omega_{\bm{K}},\lambda_{\mathrm{i}},\lambda_{\mathrm{f}})$ is observed for such frequencies of emitted photons $\omega_{\bm{K},N}$ (with integer $N$), that satisfy the condition
\begin{equation}
\bar{Q}^+/k^0N_{\mathrm{rep}}=-N,\quad \bar{Q}^+=\bar{p}^+_{\mathrm{i}}-\bar{p}^+_{\mathrm{f}}-K^+,
\label{comb2}
\end{equation}
provided that $N_{\rm rep}>1$. Here, the light-cone variable $\bar{Q}^+=(\bar{Q}^0+\bm{n}\cdot\bar{\bm{Q}})/2$ (and similarly for the remaining four-momenta of electron and photon) has been used.
It follows from Eq.~\eqref{comb2} that the probability distribution of Compton scattering ($\sim |\mathcal{A}_{\sigma}(\omega_{\bm{K}},\lambda_{\mathrm{i}},\lambda_{\mathrm{f}})|^2$) is enhanced 
by a factor of $N_{\rm rep}^2$. Meaning that the energy spectrum of radiated photons consists of a sequence of well separated peaks, which is in contrast to the Compton process induced by a single pulse 
(for which $N_{\rm rep}=1$). Note also that, in contrast to the Thomson scattering, these peaks are not \textit{exactly} equally separated from each other over the whole interval of allowed 
frequencies $(0,\omega_{\mathrm{cut}})$. When $\omega_{\bm{K}}$ approaches the cut-off value, the distribution of $\omega_{\bm{K},N}$ becomes increasingly denser. Thus, one can get the regular frequency 
combs out of the Compton radiation only within limited frequency intervals.

\section{Generalized Klein-Nishina formula and laser pulse diagnosis}
\label{method}

Eq.~\eqref{comb2} allows us to determine the peak frequencies in the energy spectrum of emitted Compton photons. After algebraic manipulations, 
we arrive at the \textit{generalized Klein-Nishina formula}, 
\begin{equation}
\omega_{\bm{K},N}^{(\rm GKN)}=\frac{(N/N_{\mathrm{osc}})\omega_{\mathrm{L}}}{\frac{p_{\mathrm{i}}\cdot n_{\bm{K}}}{p_{\mathrm{i}}\cdot n}+\frac{\nu n\cdot n_{\bm{K}}
+g_1p_{\mathrm{i},1}+g_2p_{\mathrm{i},2}}{(p_{\mathrm{i}}\cdot n)^2}+\frac{(N/N_{\mathrm{osc}})\omega_{\mathrm{L}}}{\omega_{\mathrm{cut}}}},
\label{gkn1}
\end{equation}
which for the plane wave (i.e., when $N_{\mathrm{osc}}=1$ and $g_{j}=0$) reduces to the original one \eqref{comp1}.
Here, $g_1$ and $g_2$ are defined in Eq.~\eqref{volk8},
\begin{equation}
\nu=\frac{1}{2}(\mu m_{\mathrm{e}}c)^2(\langle f_1^2\rangle + \langle f_2^2\rangle),
\label{gkn2}
\end{equation}
and (for $j=1,2$)
\begin{equation}
p_{\mathrm{i},j}=(p_{\mathrm{i}}\cdot n)(n_{\bm{K}}\cdot \varepsilon_j)-(p_{\mathrm{i}}\cdot \varepsilon_j)(n\cdot n_{\bm{K}}).
\label{gkn3}
\end{equation}
Similar to the original Klein-Nishina formula, the quantum signature of Eq.~\eqref{gkn1} is hidden in the definition of $\omega_{\mathrm{cut}}$. 
The frequencies determined by Eq.~\eqref{gkn1} mark the positions of main peaks in the Compton spectrum. Similarly, 
one can find the location of secondary peaks (if $N_{\mathrm{rep}}>2$) and zeros (if $N_{\mathrm{rep}}>1$) 
in the angular-resolved frequency distributions. It is also worth noting that the polarization-dependent terms 
appear now in the definition of the directly measurable quantity; in other words, they affect the positions of peak frequencies in the Compton spectrum. Next, we define the quantity
\begin{align}
\mathcal{N}_{\mathrm{GKN}}=&\frac{N_{\mathrm{osc}}\omega_{\bm{K}}\omega_{\mathrm{cut}}}{\omega_{\mathrm{L}} (\omega_{\mathrm{cut}}-\omega_{\bm{K}})} \nonumber \\
\times & \Bigl(\frac{p_{\mathrm{i}}\cdot n_{\bm{K}}}{p_{\mathrm{i}}\cdot n} +\frac{\nu n\cdot n_{\bm{K}}+g_1p_{\mathrm{i},1}+g_2p_{\mathrm{i},2}}{(p_{\mathrm{i}}\cdot n)^2}\Bigr),
\label{gkn4}
\end{align}
which, according to \eqref{gkn1}, acquires integer values for peak frequencies $\omega_{\bm{K},N}^{(\rm GKN)}$.

\begin{figure}
\begin{center}
\includegraphics[width=8cm]{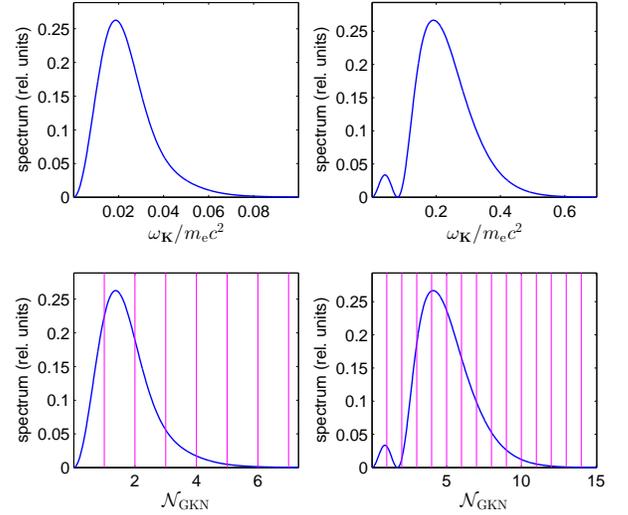}%
\end{center}
\caption{Spectra of Compton radiation resulting from the head-on collision of a linearly 
polarized laser pulse and an electron of momentum $\bm{p}_{\mathrm{i}}=-10^2m_{\mathrm{e}}c\bm{e}_z$. The laser pulse ($\mu=1$ 
and $\omega_{\mathrm{L}}=3\times 10^{-6}m_{\mathrm{e}}c^2$) propagates along the $z$ direction and it is polarized along the $x$ direction. 
These two axes define the scattering plane. The pulse has a $\sin^2$ envelope \eqref{t2} with $N_{\mathrm{osc}}=1$ and 
$N_{\mathrm{rep}}=1$. The Compton photon is emitted in the direction determined by the polar angle $\theta_{\bm{K}}=0.99\pi$ (left column) 
or $\theta_{\bm{K}}=0.995\pi$ (right column) and by the azimuthal angle $\varphi_{\bm{K}}=\pi$, 
with the polarization vector parallel to the scattering plane. In the upper row, the energy spectra of Compton radiation are plotted as functions of the frequency $\omega_{\bm{K}}$ whereas, in the lower row,
as functions of $\mathcal{N}_{\mathrm{GKN}}$. The vertical lines mark the integer values of the respective argument. 
\label{c1}}
\end{figure}

The positions of the interference peaks allow us to determine parameters of the driving laser field. In order to demonstrate this, we introduce the quantities,
\begin{align}
\eta=&\nu n\cdot n_{\bm{K}}+g_1p_{\mathrm{i},1}+g_2p_{\mathrm{i},2}, \label{keta} \\
A(\omega_{\bm{K}})=&\frac{N_{\mathrm{osc}}\omega_{\bm{K}}\omega_{\mathrm{cut}}}{\omega_{\mathrm{L}} (\omega_{\mathrm{cut}}-\omega_{\bm{K}})(p_{\mathrm{i}}\cdot n)^2},  \\
B(\omega_{\bm{K}})=&\frac{N_{\mathrm{osc}}\omega_{\bm{K}}\omega_{\mathrm{cut}}}{\omega_{\mathrm{L}} (\omega_{\mathrm{cut}}-\omega_{\bm{K}})}\frac{p_{\mathrm{i}}\cdot n_{\bm{K}}}{p_{\mathrm{i}}\cdot n}.
\label{alg1}
\end{align}
They depend on the initial electron energy and the geometry of the process. Except of $\eta$, they also depend on the frequency of 
emitted photons. The latter means that, for a given frequency distribution, $\eta$ remains constant. Moreover, only $\eta$ depends on the laser pulse parameters 
such as $\mu^2(\langle f_1^2\rangle+\langle f_2^2\rangle)$ and $\mu\langle f_j\rangle$, $j=1,2$ (note that only these parameters are necessary 
for determining the electron dressed mass). Using the above definitions, Eq.~\eqref{gkn4} can be rewritten as
\begin{equation}
A(\omega_{\bm{K}})\eta=\mathcal{N}_{\mathrm{GKN}}-B(\omega_{\bm{K}}).
\label{alg2}
\end{equation}
This equation determines $\eta$ provided that we can unambiguously prescribe an integer number $\mathcal{N}_{\mathrm{GKN}}$ to an arbitrary chosen peak. In general, this is impossible. 
Therefore, we choose instead two arbitrary, consecutive peaks from the spectrum, i.e., $\omega_{\bm{K},N}^{(\rm GKN)}$ and $\omega_{\bm{K},N+1}^{(\rm GKN)}$. 
By solving the system of two equations~\eqref{alg2} evaluated at $\omega_{\bm{K},N}^{(\rm GKN)}$ and $\omega_{\bm{K},N+1}^{(\rm GKN)}$, we obtain that
\begin{equation}
\eta=\frac{1-B(\omega_{\bm{K},N+1}^{(\rm GKN)})+B(\omega_{\bm{K},N}^{(\rm GKN)})}{A(\omega_{\bm{K},N+1}^{(\rm GKN)})-A(\omega_{\bm{K},N}^{(\rm GKN)})}.
\label{alg3}
\end{equation}
This quantity is already independent of $N$. Thus, in order to determine three independent parameters $\nu$, $g_1$, and $g_2$ present in the definition of $\eta$, 
we should repeat  this procedure for three different geometries. This finally leads to the system of three linear equations for the unknown $\nu$, $g_1$, and $g_2$.

\section{Illustration of the proposed diagnostic method}

\begin{figure}
\begin{center}
\includegraphics[width=8cm]{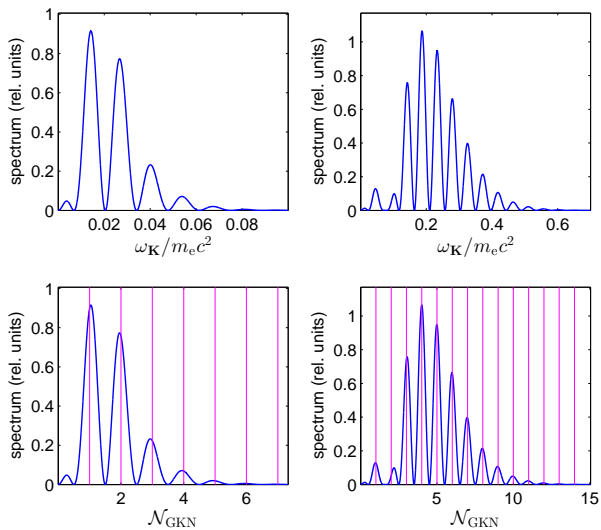}%
\end{center}
\caption{The same as in Fig.~\ref{c1} but for $N_{\rm rep}=2$. We clearly see that the peaks appear for integer values of $\mathcal{N}_{\mathrm{GKN}}$.
\label{c2}}
\end{figure}

In the following, we shall consider the linearly polarized laser pulse such that, for $0\leqslant \phi=k\cdot x\leqslant 2\pi$, the four-vector potential has 
the form $A(\phi)=A_0 \varepsilon f(\phi)$ [i.e., in Eq.~\eqref{com1}, we take $\varepsilon=\varepsilon_1$, $f(\phi)=f_1(\phi)$, $f_2(\phi)=0$] and the electric field vector 
is $\bm{\mathcal{E}}(\phi)=-\omega A_0 \bm{\varepsilon} f^{\prime}(\phi)$. The shape function $f(\phi)$ is defined via its derivative,
\begin{equation}
f'(\phi)=\begin{cases} 0, & \phi <0, \cr
                 N^{\prime}_f\sin^2\bigl(N_{\mathrm{rep}}\frac{\phi}{2}\bigr)\sin(N_{\mathrm{rep}}N_{\mathrm{osc}}\phi), & 0\leqslant\phi\leqslant 2\pi,\cr
								0, & \phi > 2\pi,
			  \end{cases}
\label{t2}
\end{equation}
where we assume that $f(0)=0$. The pulse described by this shape function consists of $N_{\rm rep}$ identical subpulses, with $N_{\rm osc}$ cycles each. In other words,
it can be treated as a finite train of $N_{\rm rep}$ pulses. Regardless of this interpretation, it is justified to apply the theory derived in Ref.~\cite{KKcompton} for the process driven by a single pulse.
Since the modulated pulse~\eqref{t2} lasts for time $T_{\mathrm{p}}$ we can define its fundamental, $\omega=2\pi/T_{\mathrm{p}}$, and its central frequency, 
$\omega_{\mathrm{L}}=N_{\mathrm{rep}}N_{\mathrm{osc}}\omega$. The latter is supposed to be fixed and equal to $\omega_{\mathrm{L}}=3\times 10^{-6}m_{\mathrm{e}}c^2$ in all calculations performed below.  
In the following, we assume also that, in Eq.~\eqref{t2}, we have $N^{\prime}_f=N_{\mathrm{rep}}N_{\mathrm{osc}}$. This guarantees that the time-averaged intensity 
carried out by the laser field is independent of $N_{\mathrm{rep}}$ and $N_{\mathrm{osc}}$, as for this particular choice of $N^\prime_f$ the amplitude 
of the electric field scales as $\mu\omega_{\mathrm{L}}$. 

\begin{figure}
\begin{center}
\includegraphics[width=8cm]{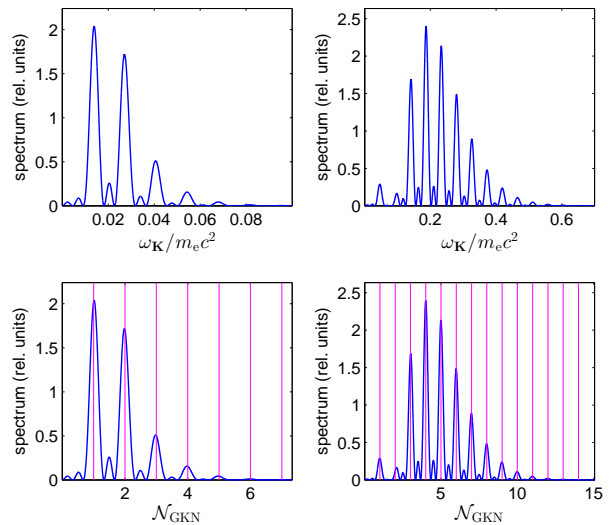}%
\end{center}
\caption{The same as in Fig.~\ref{c1} but for $N_{\rm rep}=3$. We clearly see that the peaks appear for integer values of $\mathcal{N}_{\mathrm{GKN}}$.
\label{c3}}
\end{figure}

To illustrate the diagnostic method described in the previous Section, we choose the extremely short laser pulse consisting of only one cycle, i.e., in Eq.~\eqref{t2}, we take
$N_{\mathrm{osc}}=1$ and $N_{\mathrm{rep}}=1$. In Fig.~\ref{c1}, we present the energy spectrum of radiation generated by such a pulse as functions of either the Compton photon frequency $\omega_{\bm{K}}$ 
or the parameter $\mathcal{N}_{\mathrm{GKN}}$, for two different kinematics. For the polar angle $\theta_{\bm{K}}=0.99\pi$ (left column), one can hardly attribute an integer value
of $\mathcal{N}_{\mathrm{GKN}}$ to the broad peak which appears in the spectrum. On the other hand, for $\theta_{\bm{K}}=0.995\pi$ (right column), one can prescribe integer values $\mathcal{N}_{\mathrm{GKN}}=1$ and 
$\mathcal{N}_{\mathrm{GKN}}=4$ to both peaks, even though there are no additional peaks in between them. Let us also note that, if one tried to associate to both peaks the consecutive 
integers $\mathcal{N}_{\mathrm{GKN}}=1$ and $\mathcal{N}_{\mathrm{GKN}}=2$, one would obtain the incorrect estimation of the laser pulse characteristics $\langle f\rangle$ and $\langle f^2\rangle$.

\begin{figure}
\begin{center}
\includegraphics[width=8cm]{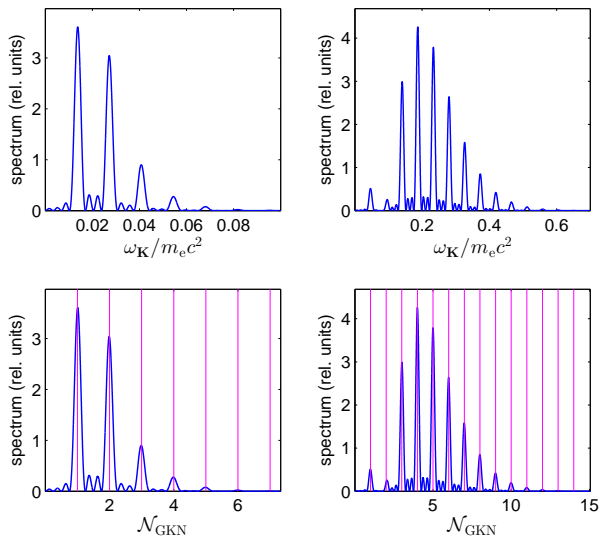}%
\end{center}
\caption{The same as in Fig.~\ref{c1} but for $N_{\rm rep}=4$. We clearly see that the peaks appear for integer values of $\mathcal{N}_{\mathrm{GKN}}$.
\label{c4}}
\end{figure}

The situation changes if we apply such a pulse twice, meaning that $N_{\mathrm{osc}}=1$ and $N_{\mathrm{rep}}=2$ in Eq.~\eqref{t2}. Now, as it is demonstrated in Fig.~\ref{c2} for both polar 
angles, we observe peaks in the spectrum to which one can prescribe unambiguously integer values of $\mathcal{N}_{\mathrm{GKN}}$. This is particularly evident for the most pronounced peaks. 
Next, choosing for both cases two adjacent peaks from these spectra, we acquire from the numerical (or experimental) data the corresponding Compton frequencies (without knowing what the integer 
values of $\mathcal{N}_{\mathrm{GKN}}$ are prescribed to them). Applying the method described in Sec.~\ref{method} one can estimate now the laser pulse characteristics.
Namely, for the considered shape function~\eqref{t2}, they are very close to the exact values $\langle f\rangle =-0.375$ and $\langle f^2\rangle = 0.1328125$.
Note that, for larger $N_{\mathrm{rep}}$, the peaks in the Compton spectrum practically do not change their positions and they become sharper (see, Figs.~\ref{c3} and \ref{c4}). Thus, for larger $N_{\rm rep}$, the determination 
of the laser pulse properties can be made with an increased precision.

\section{Conclusions}

The aim of this paper was to discuss the concept of the electron mass dressing by a short and intense laser pulse. We have shown that this concept seems to be not as fundamental 
as for the monochromatic plane wave (in general, for the polychromatic plane waves with commensurate frequencies), as it does not follow directly from the Floquet-Bloch analysis 
of the Volkov solution. However, with some modifications, it is possible to redefine the electron momentum dressing such that it is on the mass shell with the dressed mass, 
independent of the asymptotic electron momentum. If such a modification is accepted, then the analysis of the frequency combs in the Compton scattering allows one to measure 
the dressed electron mass for arbitrarily short (even one-cycle) and intense laser pulses. This is achieved by determining characteristics of the laser pulse. 
From the practical point of view, the procedure described above provides a very precise method of diagnosis of extremely intense pulses in cases when the conventional 
methods fail. Note also that this method is not only limited to the Compton or Thomson scattering, but can be extended to other quantum processes as well (for instance, to the Breit-Wheeler 
process~\cite{KK3} or to ionization of atoms and molecules~\cite{KKcombs2015,CKK}).


\begin{thebibliography}{99}

\bibitem{RitusNikishov}
V. I. Ritus and A. I. Nikishov, \textit{Quantum Electrodynamics Phenomena in the Intense Field}, Trudy FIAN \textbf{111}, 5 (1979).

\bibitem{Ritus}
V. I. Ritus, J. Sov. Laser Res. {\bf 6}, 497 (1985).

\bibitem{RitusNikishov2}
V. I. Ritus and A. I. Nikishov, \textit{Problems of Intense Field Quantum Electrodynamics}, Trudy FIAN \textbf{168}, 5 (1986).

\bibitem{EKK2009} 
F. Ehlotzky, K. Krajewska, and J. Z. Kami\'nski, Rep. Prog. Phys. \textbf{72}, 046401 (2009).

\bibitem{Rev1}
R. Ruffini, G. Vereshchagin, and S. S. Xue, Phys. Rep. \textbf{487}, 1 (2010).

\bibitem{PMHK2012} 
A. Di Piazza, C. M\"uller, K. Z. Hatsagortsyan, and C. H. Keitel, Rev. Mod. Phys. {\bf 84}, 1177 (2012).

\bibitem{Ras}
V. N. Nedoreshta, S. P. Roshchupkin, and A. I. Voroshilo, Laser Phys. {\bf 23}, 055301 (2013).

\bibitem{Narozhny}
N. B. Narozhny and A. M. Fedotov, Contemp. Phys. \textbf{56}, 249 (2015).

\bibitem{KleinNishina}
O. Klein and Y. Nishina, Z. Physik {\bf 52}, 853 (1929).

\bibitem{BrownKibble}
L. S. Brown and T. W. B. Kibble, Phys. Rev. \textbf{133}, A705 (1964).

\bibitem{SKscale1}
T. Heinzl, D. Seipt, and B. K\"ampfer, Phys. Rev. A {\bf 81}, 022125 (2010).

\bibitem{SKscale2}
D. Seipt and B. K\"ampfer, Phys. Rev. A {\bf 83}, 022101 (2011).

\bibitem{SKscale3}
D. Seipt and B. K\"ampfer, Phys. Rev. ST Accel. Beams {\bf 14}, 040704 (2011).

\bibitem{KKscaling}
K. Krajewska and J. Z. Kami\'nski, Phys. Rev. A {\bf 90}, 052117 (2014).

\bibitem{Volkov}
D. M. Volkov, Z. Phys. {\bf 94}, 250 (1935).

\bibitem{Lee2010}
K. Lee, S. Y. Chung, S. H. Park, Y. U. Jeong, and D. Kim, Europhys. Lett. {\bf 89}, 64006 (2010).
 
\bibitem{KKcompton}
K. Krajewska and J. Z. Kami\'nski, Phys. Rev. A {\bf 85}, 062102 (2012).

\bibitem{KKbw}
K. Krajewska and J. Z. Kami\'nski, Phys. Rev. A {\bf 86}, 052104 (2012).

\bibitem{KKgkn}
K. Krajewska, F. Cajiao V\'elez, and J. Z. Kami\'nski, Phys. Rev. A {\bf 91}, 062106 (2015).

\bibitem{Krec1}
A. Jaro\'n, J. Z. Kami\'nski, and F. Ehlotzky, Phys. Rev. A \textbf{61}, 023404 (2000).

\bibitem{Krec2}
A. Jaro\'n, J. Z. Kami\'nski, and F. Ehlotzky, Phys. Rev. A \textbf{63}, 055401 (2001).

\bibitem{Carstenrec}
C. M\"uller, A. B. Voitkiv, and B. Najjari, J. Phys. B \textbf{42}, 221001 (2009).

\bibitem{Har}
O. Har-Shemesh and A. Di Piazza, Opt. Lett. {\bf 37}, 1352 (2012).

\bibitem{Kalashnikov}
M. Kalashnikov, A. Andreev, K. Ivanov, A. Galkin, V. Korobkin, M. Romanovsky, O. Shiryaev, M. Schnuerer, J. Braenzen, and V. Trofimov, Laser Part. Beams \textbf{33}, 361 (2015).

\bibitem{KKcombs2015}
K. Krajewska and J. Z. Kami\'nski, Phys. Lett. A (submitted).

\bibitem{Brad}
I. Ghebregziabher, B. A. Shadwick, and D. Umstadter, Phys. Rev. ST Accel. Beams {\bf 16}, 030705 (2013).

\bibitem{Terzic}
B. Terzi\'c, K. Deitrick, A. S. Hofler, and G. A. Krafft, Phys. Rev. Lett. {\bf 112}, 074801 (2014).

\bibitem{Seipt2015}
D. Seipt, S. G. Rykovanov, A. Surzhykov, and S. Fritzsche, Phys. Rev. A {\bf 91}, 033402 (2015).

\bibitem{HHIM}
C. Harvey, T. Heinzl, A. Ilderton, and M. Marklund, Phys. Rev. Lett. {\bf 109}, 100402 (2012).

\bibitem{Young}
T. Young, Phil. Trans. R. Soc. Lond. \textbf{94}, 1 (1804).

\bibitem{freqcombs1}
J. L. Hall, Rev. Mod. Phys. \textbf{78}, 1279 (2006).

\bibitem{freqcombs2}
T. W. H\"ansch, Rev. Mod. Phys. \textbf{78}, 1297 (2006).

\bibitem{freqcombs3}
H. S. Margolis, J. Phys. B: At. Mol. Opt. Phys. \textbf{42}, 154017 (2009).

\bibitem{AgostiniMauro}
P. Agostini and L. F. DiMauro, Rep. Prog. Phys. \textbf{67}, 813 (2004).

\bibitem{FT}
Gy. Farkas and Cs. T\'oth, Phys. Lett. A \textbf{168}, 447 (1992).

\bibitem{KrauszIvanov}
F. Krausz and M. Ivanov, Rev. Mod. Phys. \textbf{81}, 163 (2009).

\bibitem{KK1}
K. Krajewska and J. Z. Kami\'nski, Laser Phys. Lett. \textbf{11}, 035301 (2014).

\bibitem{KK2}
K. Krajewska, M. Twardy and J. Z. Kami\'nski, Phys. Rev. A \textbf{89}, 052123 (2014).

\bibitem{KK3}
K. Krajewska and J. Z. Kami\'nski, Phys. Rev. A \textbf{90}, 052108 (2014).

\bibitem{CKK}
F. Cajiao V\'elez, K. Krajewska, and J. Z. Kami\'nski, Phys. Rev. A {\bf 91}, 053417 (2015).

\end{thebibliography}
\end{document}